\documentclass[lettersize,journal]{IEEEtran}

\usepackage{orcidlink}
\usepackage{amsmath,amssymb,amsthm,bm,graphicx,booktabs,array,cite,url}
\usepackage[caption=false,font=footnotesize]{subfig}

\newtheorem{theorem}{Theorem}
\newtheorem{proposition}{Proposition}
\newtheorem{lemma}{Lemma}
\newtheorem{corollary}{Corollary}
\newcommand{\CN}{\mathcal{CN}}
\newcommand{\E}{\mathbb{E}}
\newcommand{\Prob}{\mathbb{P}}
\newcommand{\C}{\mathcal{C}}
\newcommand{\tr}{\operatorname{tr}}
\newcommand{\range}{\operatorname{range}}

\newcommand{\FA}{\mathrm{FA}}
\newcommand{\FR}{\mathrm{FR}}
\newcommand{\lo}{\underline}
\newcommand{\hi}{\overline}
\newcommand{\norm}[1]{\left\lVert#1\right\rVert}

\newcommand{\MCTotal}{4150000}
\newcommand{\MCEnroll}{3650000}
\newcommand{\CalibrationEll}{0.96167}
\newcommand{\CalibrationG}{413.148}
\newcommand{\CalibrationXi}{0.42441}
\newcommand{\CalibrationEta}{0.70205}
\newcommand{\FRBound}{0.009984}
\newcommand{\LegitFloor}{0.35760}
\newcommand{\AttackFloor}{0.01065}

\begin{document}
	\title{Near-Field Physical-Layer Authentication Under Noisy Enrollment  }
	\author{Parsa~Rajabi~\orcidlink{0009-0005-4645-9444},
		Mohammad~Reza~Abedi~\orcidlink{0000-0003-4114-1339}~\IEEEmembership{Member,~IEEE},
		Paeiz~Azmi~\orcidlink{0000-0001-9736-3462}~\IEEEmembership{Senior~Member,~IEEE},
		Nader~Mokari~\orcidlink{0000-0001-5364-8888}~\IEEEmembership{Senior~Member,~IEEE}
		\thanks{This work has been submitted to the IEEE for possible publication.
			Copyright may be transferred without notice, after which this version may
			no longer be accessible.}
		\thanks{P. Rajabi, M. R. Abedi, P. Azmi, and N. Mokari are with the Department of Electrical and Computer Engineering, Tarbiat Modares University, Tehran, 14115-111, Iran (e-mail: parsa\_rajabi@modares.ac.ir; Mohammadreza\_abedi@modares.ac.ir; pazmi@modares.ac.ir; nader.mokari@modares.ac.ir).}
	}
	\markboth{}%
	{ }
	\maketitle
	\begin{abstract}
		In near-field physical-layer authentication (PLA), the receiver estimates a spatial
		reference from noisy enrollment pilots. With a fixed residual-energy
		threshold, even a small reference error can cause the false rejection (FR)
		probability to approach one as the received energy increases. We derive the
		distribution of the overlap between the true and enrolled spatial directions
		and use it to calibrate a test based on the aligned energy and the
		residual-to-aligned energy ratio. The calibration bounds FR for arbitrary regional gains and all legitimate received energies above a
		prescribed minimum. For a fixed enrolled reference, we reduce the problem
		of finding the maximum acceptance probability of a multi-antenna attacker
		to a one-dimensional search. We also derive a pre-enrollment upper bound
		that holds when the attacker chooses its precoder after observing the
		enrolled reference. For a single region and a fixed angular threshold, the
		enrollment energy sufficient to preserve angular separation grows inversely
		with the square of the available angular margin. In the numerical study with $128$ receive elements at $28$~GHz,
		the FR upper bound was $0.9984\%$, including the enrollment
		outage allowance. At a regional phase perturbation of $20^\circ$,
		four regions provided the best trade-off among the tested partitions:
		its simulated FR was $0.1442\%$, while its false acceptance (FA) at $9$~m was
		$1.084\%$ under the fixed transmit-energy setting.
	\end{abstract}
	
	\section{Introduction}
	
	\IEEEPARstart{A}{n} authentication receiver compares a newly
	observed physical signature with a trusted reference. In a
	large-aperture array, spherical-wave curvature provides
	range-dependent information in addition to bearing. This can
	help distinguish an impersonator at the legitimate bearing.
	However, the resulting authentication performance depends on
	reference-acquisition errors, receiver noise, unknown channel
	gains, and the attacker's transmission strategy. A nonzero
	mismatch between noiseless signatures alone does not determine
	the probability that a receiver accepts an attack.
	
	Recent physical-layer authentication (PLA) methods exploit
	channel, spatial, and device-dependent features. Multiple
	channel-based features are combined for authentication
	in~\cite{xie2021features}, while beam-pattern features are
	investigated in millimeter-wave multiple-input multiple-output
	(MIMO) systems in~\cite{zhang2025beam}. Spatial channel sparsity
	is exploited for authentication in aerial-to-ground MIMO
	links~\cite{teng2024sparsity}. Array-pattern features are also
	used in a distributed authentication
	framework~\cite{zhang2025distributed}, and array-error features
	are combined with two-beam transmission to improve
	authentication reliability~\cite{zhang2025reliable}. A related
	device-fingerprinting approach uses channel reciprocity for
	channel-robust radio frequency fingerprint
	extraction~\cite{dong2026reciprocity}. 
	Collaborative PLA has also been studied, with incentives
	to encourage reliable nodes to participate in the
	authentication process~\cite{zhou2026collaborative}.
	These studies illustrate
	the importance of both discriminative features and their
	sensitivity to propagation and hardware effects.
	
	Reference adaptation is another important issue in practical
	PLA. A tracking-based method updates beamspace signatures to
	accommodate mobility~\cite{afeef2024tracking}. A model-driven
	learning approach combines hypothesis testing with a
	lightweight neural network for mobile Wi-Fi
	devices~\cite{guo2026model}, while channel extrapolation and
	generative artificial intelligence are used to adapt
	authentication fingerprints to dynamic
	environments~\cite{cheng2026apeg}. Complex-valued neural networks
	have also been investigated for authentication against pilot
	contamination and clone attacks~\cite{zeng2026intelligent}.
	Our setting addresses a different source of uncertainty:
	a spatial reference estimated from a finite trusted enrollment.
	We retain this noisy reference during verification and analyze
	its effect without learning a decision rule. Pilot contamination
	during trusted enrollment is outside the guarantees developed
	here.
	
	Near-field propagation provides additional spatial information
	for PLA. 
	Recent work surveys spherical-wave channel models and
	estimation methods for near-field systems~\cite{long2026nearfield}.
	Antenna placement has also been optimized to improve the
	achievable rate of near-field links with movable
	antennas~\cite{liu2026movable}.
	Visibility-region signatures are used for authentication
	in extremely large-scale MIMO
	systems~\cite{demirci2025,wang2025}. Related work uses
	visibility-region information to identify environmental
	surfaces~\cite{surface2026}; this is a surface-identification
	task rather than transmitter authentication. Near-field
	backscatter authentication also exploits geometry-dependent
	delay--curvature signatures~\cite{uledi2026}. These results
	motivate the use of aperture-dependent signatures, but a
	receiver must also account for uncertainty in their acquisition.
	We represent unknown regional complex gains by a legitimate
	subspace and estimate its regional directions from independent
	noisy enrollment observations.
	
	Near-field security has also been studied against
	eavesdropping. Curved beams can steer energy away from
	regions where an eavesdropper may be
	located~\cite{liu2026caustic}, while deceptive signals
	can help protect communication against moving
	eavesdroppers~\cite{he2026deception}. These studies address
	information leakage, whereas our receiver decides whether
	to accept a received pilot as legitimate.
	
	Attack-aware analysis is necessary because an informed
	transmitter can deliberately alter the observed signature.
	Power, phase, and position attacks are analyzed for distributed
	single-input multiple-output (SIMO) authentication
	in~\cite{forssell2020}. Cooperative impersonation strategies
	under energy and information constraints are studied
	in~\cite{pourafzal2025}. Angle-of-arrival (AoA) authentication
	has also been analyzed against single- and multiple-antenna
	impersonators~\cite{pham2026aoa}. For near-field authentication,
	the analysis in~\cite{hassani2026} minimizes signature
	mean-square error (MSE) and characterizes scalar and
	multi-antenna matching conditions. We use this
	signature-matching formulation as a physical attack reference.
	Our attack objective is instead the probability of passing the
	specified noisy-reference receiver test under a transmit-energy
	budget.
	
	Reconfigurable intelligent surfaces (RISs) provide additional
	ways to construct authentication features and challenges.
	Cascaded-channel signatures are used for PLA in RIS-assisted
	communication systems~\cite{zhang2025cascaded}.
	Challenge--response authentication is developed between ambient
	backscatter devices~\cite{zhang2026backscatter}, while
	receiver-controlled RIS configurations are optimized to balance
	communication and authentication performance
	in~\cite{guglielmi2026challenge}. 
	For orthogonal frequency-division multiplexing (OFDM)
	systems, channel randomness tests have been proposed to
	determine when wireless challenge--response PLA can be
	used~\cite{liu2026ofdmpla}.
	Our setting uses a stored noisy
	spatial reference and grants the attacker knowledge of its
	realized value. The attacker may then optimize its precoder
	under a transmit-energy constraint. The security analysis
	therefore conditions on the enrolled template before optimizing
	the attack and subsequently derives a certificate that also
	applies before enrollment.
	
	The focus of this work is the joint treatment of finite
	enrollment, regional gain uncertainty, and physically constrained
	attacks that adapt to the realized template. 
	For a given enrolled reference, the objective is to maximize the probability
	that a spoofed pilot passes the receiver's test over all attacker precoders
	that satisfy the transmit-energy constraint. We share our code in~\cite{rajabi2026enrollmentcode}.
	The contributions are as follows.
	\begin{itemize}
		\item We propagate finite enrollment through the verifier using
		a doubly noncentral beta overlap law. This yields exact
		fixed-attack averaging identities, verification-energy limits,
		and a failure result for fixed residual-energy ceilings.
		
		\item We derive a constructive false rejection (FR) guarantee that
		is uniform over all legitimate verification energies above a
		prescribed minimum, with no upper energy bound, and over
		arbitrary regional complex gains, conditional on a
		high-probability enrollment event.
		
		\item We characterize the template-aware attack optimum through
		its attainable aligned and residual energies. We also derive
		a convex two-parameter Chernoff certificate with an explicit
		Schur-complement formulation for arbitrary legitimate-subspace
		rank.
		
		\item We link enrollment energy to adversarial subspace
		separation. The resulting pre-enrollment certificate remains
		valid when the attacker adapts to the realized noisy template,
		and the sufficient enrollment-energy budget for angular
		certification follows an inverse-square law as the available
		angular margin vanishes.
	\end{itemize}

	\noindent\textit{Notation:}
	Bold lowercase and uppercase letters denote vectors and matrices,
	respectively. The superscript $(\cdot)^H$ denotes the Hermitian
	transpose, $\|\cdot\|$ denotes the Euclidean norm, and $\bm I_d$
	is the $d\times d$ identity matrix. The notation
	$\bm X\succeq0$ means that $\bm X$ is Hermitian positive
	semidefinite. We write $\operatorname{tr}(\bm X)$,
	$\lambda_{\max}(\bm X)$, and $\operatorname{range}(\bm X)$
	for the trace, largest eigenvalue, and column space of a matrix.
	The distribution $\mathcal{CN}(\bm m,\bm\Sigma)$ is a proper
	complex Gaussian distribution with mean $\bm m$ and covariance
	$\bm\Sigma$. For a random variable $X$, $F_X$ and $F_X^{-1}$
	denote its cumulative distribution function (CDF) and quantile
	function, respectively. The notation $\chi'^2_d(\nu)$ denotes
	a noncentral chi-square distribution with $d$ degrees of freedom
	and noncentrality parameter $\nu$. The notation $(x)_+=\max\{x,0\}$ denotes the positive part of a real number. All logarithms are natural,
	and all angles in analytical expressions are in radians.

	\section{System Model}
We model the near-field channel, the acquisition of a trusted reference,
and the received pilot used for verification.
\begin{figure}[t]
	\centering
	\includegraphics[width=\columnwidth]{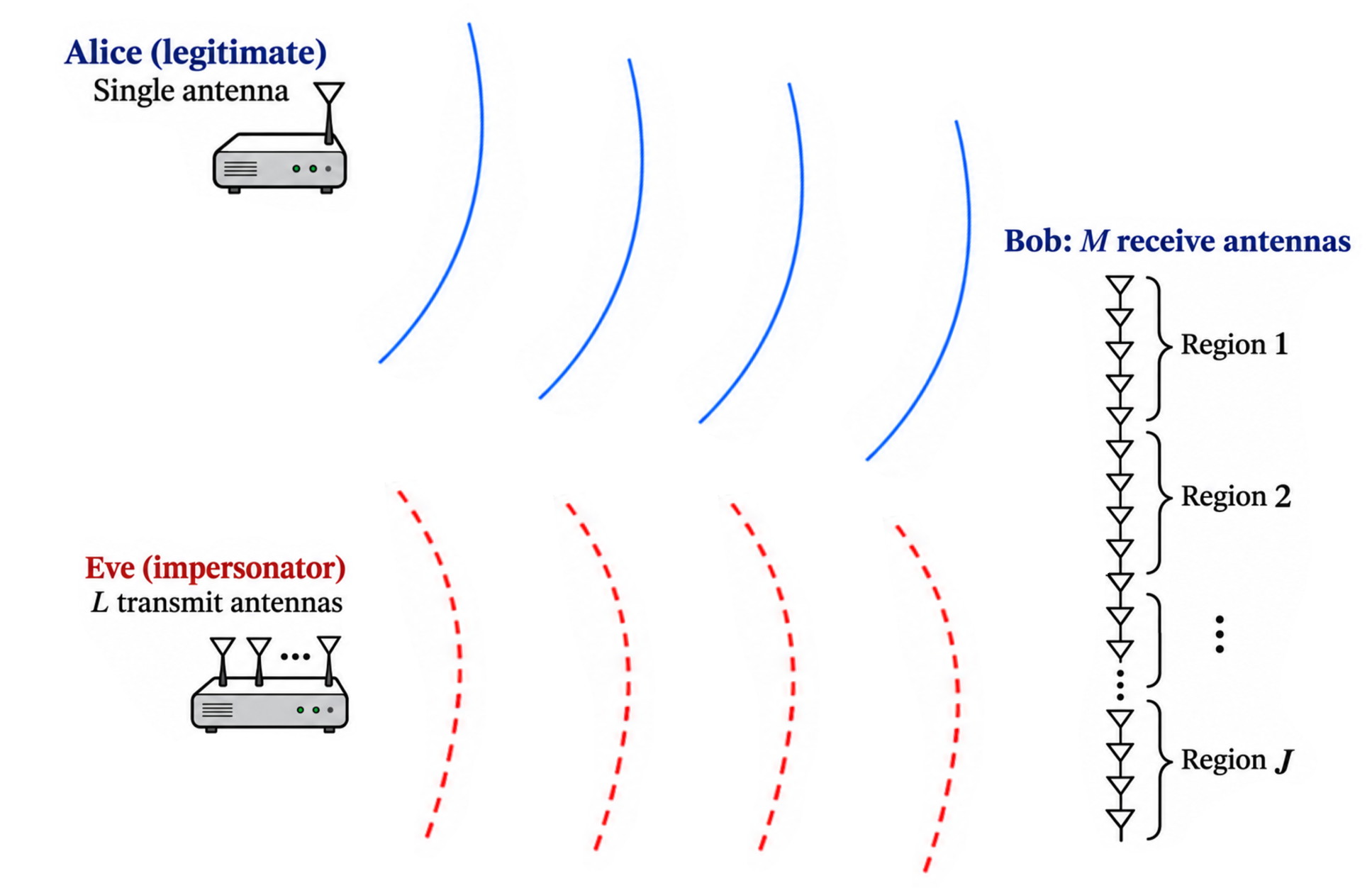}
	\caption{Geometry of the near-field verification model.}
	\label{fig:system-model}
\end{figure}
As shown in Fig.~\ref{fig:system-model}, Bob has $M$ calibrated
receive elements on a linear array. The coordinate $x_m$ is
measured along the array axis from its center. For any source,
the range $r$ is measured from Bob's array center, and the angle
$\theta\in(-\pi/2,\pi/2)$ is measured from broadside, with positive
angles toward the positive $x$ direction. We use
$(r_A,\theta_A)$ for Alice's transmit antenna and
$(r_E,\theta_E)$ for the center of Eve's transmit array.
For a narrowband line-of-sight source at $(r,\theta)$, the
distance and channel coefficient to element $m$ are
\begin{align}
	d_m(r,\theta)&=\sqrt{r^2+x_m^2-2rx_m\sin\theta},\label{eq:distance}\\
	h_m(r,\theta)&=\frac{b_m R_0}{\sqrt M\,d_m(r,\theta)}
	e^{-\mathrm j\kappa[d_m(r,\theta)-r]},\quad \kappa=2\pi/\lambda .\label{eq:channel}
\end{align}
The carrier wavelength is $\lambda$, $R_0$ is a reference range, and
$b_m$ is a fixed factor that accounts for element response and visibility.
We absorb the common propagation phase into a complex gain.
Equation~\eqref{eq:channel} includes the spherical-wave phase and
the $1/d_m$ amplitude variation across the array. It omits mutual
coupling and polarization. We use exact
distances throughout the physical-channel experiments, so no Fresnel
expansion is needed. References~\cite{bjornson2021,carvalho2020}
discuss near-field propagation and spatial nonstationarity.
The channel vector from a single transmit antenna at
$(r,\theta)$ to Bob is $\bm h(r,\theta)
= [h_1(r,\theta),\ldots,h_M(r,\theta)]^T
\in \mathbb C^M$.

We divide the active aperture into $J$ disjoint regions $\mathcal I_j$,
where $m_j=|\mathcal I_j|\ge2$ and $\sum_j m_j=M$. We assume that
Alice's response has nonzero norm in each region. We normalize each
regional response and fill the entries outside that region with zeros
to obtain $\bm u_j\in\mathbb C^M$. The resulting basis is
$\bm U=[\bm u_1,\ldots,\bm u_J]$, with $\bm U^H\bm U=\bm I_J$.
Alice's mean is $\bm\mu_A=\bm U\bm\beta$.
Each regional gain $\beta_j$ can take any complex value. Let
$E_j=|\beta_j|^2$ and impose $E=\sum_jE_j\ge E_{\min}>0$.
The regions and their directions are fixed across enrollment and
verification. The complex gains may change, as with blockwise phase
drift. The model assumes that motion and other effects leave the
direction within each region unchanged.

During trusted enrollment, Bob matches the received signal to the
pilots and divides by the known standard deviation of the noise
after filtering. The observation in region $j$ is
\begin{equation}
	\bm z_j=\sqrt{\tau_j}\,\bm u_{j,\mathcal I_j}+\bm\epsilon_j,
	\label{eq:enrollment}
\end{equation}
where $\bm\epsilon_j\sim\CN(\bm0,\bm I_{m_j})$ is the normalized
enrollment noise. Bob estimates the regional direction as
\begin{equation}
	\hat{\bm u}_{j,\mathcal I_j}
	=\frac{\bm z_j}{\norm{\bm z_j}}.
	\label{eq:enrollment_direction}
\end{equation}

The overlap in region $j$ is
$C_j=|\hat{\bm u}_{j,\mathcal I_j}^{H}\bm u_{j,\mathcal I_j}|^2$.
The direction $\bm u_j$ includes any enrollment phase.
Enrollment noise is independent across regions. Padding each
estimated direction with zeros and stacking the columns gives
$\hat{\bm U}$, whose columns are orthonormal. The aligned and residual
dimensions are $q=J$ and $k=M-J$, respectively. The corresponding
orthogonal projectors are $\hat{\bm P}=\hat{\bm U}\hat{\bm U}^H$ and
$\hat{\bm Q}=\bm I_M-\hat{\bm P}$.

The dimensionless energy $\tau_j$ is the received enrollment
energy divided by the noise variance. Its relation to a pilot
count depends on the received quality of the pilots. If Bob
coherently averages $K$ equal-quality pilots, then
$\tau_j=K\gamma_j$, where $\gamma_j$ is the received
signal-to-noise ratio (SNR) per pilot in region $j$.
Calibration uses trusted lower bounds $\lo\tau_j$ and $E_{\min}$.
The total lower bound on the received enrollment energy is
$T=\sum_{j=1}^{J}\underline{\tau}_j$.
We assume that the acquisition procedure establishes these bounds
and the initial trust independently of the simulations and of any
unverified claimant.

At verification, Bob observes
\begin{equation}
	\bm y=\bm\mu+\bm n,
	\label{eq:online}
\end{equation}
where $\bm n\sim\CN(\bm0,\bm I_M)$ is independent of the enrollment
noise and of the transmitted signal. The hypothesis
$\mathcal H_0$ denotes a legitimate transmission and $\mathcal H_1$
denotes impersonation. The mean under the two hypotheses is
\begin{equation}
	\bm\mu=
	\begin{cases}
		\bm U\bm\beta, & \mathcal H_0,\\
		\bm H_E\bm w, & \mathcal H_1,
	\end{cases}
	\label{eq:mean_vector}
\end{equation}
where $\bm H_E\in\mathbb C^{M\times L}$ is the channel from Eve
to Bob. Let $(r_{E,l},\theta_{E,l})$ denote the position of
Eve's $l$th transmit antenna, measured from Bob's array center
using the same range and angle convention as in
\eqref{eq:distance}. The $l$th column of $\bm H_E$ is
$\bm h(r_{E,l},\theta_{E,l})$, for $l=1,\ldots,L$.
We model coherent pilot
spoofing with a normalized transmit-energy budget $P$, so the
precoder must satisfy $\norm{\bm w}^2\le P$. The channel matrix
$\bm H_E$ includes propagation attenuation.

Eve may choose $\bm w$ using the realized template $\hat{\bm U}$,
the channel $\bm H_E$, and Bob's decision rule. The future receiver
noise $\bm n$ is independent and unknown when she makes that choice.
In the conditional analysis we hold $\hat{\bm U}$ fixed. In the
population analysis we average over independent trusted enrollments.
The fixed-attack mixture applies when the mean is independent of
enrollment. For a template-aware attack, we optimize conditionally
before averaging.

Rejecting a legitimate transmission counts as an FR. Accepting an attack counts as a false acceptance (FA).
Their probabilities are $P_{\FR}=\Prob(\text{reject}\mid\mathcal H_0)$
and $P_{\FA}=\Prob(\text{accept}\mid\mathcal H_1)$, respectively.
The guarantees assume known noise scale and covariance, trusted
enrollment without pilot contamination, and narrowband propagation.
Temporal replay defenses are outside the model. A receiver with a
misspecified noise model needs recalibration. A constant false alarm
rate (CFAR) with arbitrary unknown noise covariance lies outside
the guarantee.

\section{Problem Formulation}
We derive the receiver test from the enrollment overlap and calibrate it
over the allowed legitimate gains. We then characterize the attack
probabilities, derive bounds using the array geometry, and describe
how to implement the receiver.

\subsection{Enrollment Laws and a Gain-Consistent Test}
\subsubsection{Directional Uncertainty}
Consider one region with dimension $m\ge2$. Let
$\bm u,\bm v\in\mathbb C^m$ be deterministic unit vectors, let $\tau\ge0$,
and take $\bm\epsilon\sim\CN(\bm0,\bm I_m)$. The direction $\bm v$ is fixed
independently of enrollment. Its squared overlap with $\bm u$ is
$c=|\bm u^H\bm v|^2$. Write $\pi_a(i)=e^{-a}a^i/i!$ for the Poisson
weights, with $\pi_0(0)=1$. Let $I_z(a,b)$ denote the regularized
incomplete beta function.

\begin{proposition}[Exact enrollment overlap]\label{pr:overlap}
	Let $\hat{\bm u}=(\sqrt\tau\bm u+\bm\epsilon)/\norm{\sqrt\tau\bm u+\bm\epsilon}$.
	The overlap with $\bm v$ has distribution
	\begin{equation}
		Z=|\hat{\bm u}^H\bm v|^2\sim
		\operatorname{B}_{\rm nc}(1,m-1;2\tau c,2\tau(1-c)),\label{eq:beta}
	\end{equation}
	where the last two parameters give the noncentralities of the
	energy along $\bm v$ and of the energy in its orthogonal
	complement. The CDF is
	\begin{equation}
		F_Z(z)=\sum_{i,j\ge0}\pi_{\tau c}(i)\pi_{\tau(1-c)}(j)
		I_z(1+i,m-1+j).\label{eq:betacdf}
	\end{equation}
	The legitimate overlap $C=|\hat{\bm u}^H\bm u|^2$ is the singly
	noncentral case with $c=1$.
\end{proposition}
\begin{proof}
	Project $\bm z=\sqrt\tau\bm u+\bm\epsilon$ onto $\bm v$ and onto its
	orthogonal complement. The complex Gaussian projections are
	independent. Their scaled energies satisfy
	\begin{equation}
		\begin{aligned}
			2|\bm v^H\bm z|^2&\sim\chi'^2_2(2\tau c),\\
			2\norm{(\bm I-\bm v\bm v^H)\bm z}^2&\sim\chi'^2_{2m-2}(2\tau(1-c)).
		\end{aligned}\label{eq:enrollchi}
	\end{equation}
	Dividing the first energy by the sum gives~\eqref{eq:beta}.
	For the CDF, we condition on the independent Poisson counts of
	the two noncentral chi-square variables. This gives
	\eqref{eq:betacdf} by the beta-mixture representation
	in~\cite{orsi2017}. We use this distribution below to account for
	enrollment noise in the acceptance probabilities.
\end{proof}

All mixture weights are nonnegative. When we truncate the Poisson
supports, the error in a CDF or in the expectation of a
$[0,1]$-valued function is at most the sum of the two omitted tail
masses. To evaluate the series, we sum over the numerator count
through a noncentral $F$ distribution. For a given denominator
count $j$, set $b=m-1+j$. The argument is $bz/(1-z)$, with degrees
of freedom $(2,2b)$ and noncentrality $2\tau c$. This leaves a
single sum over $j$. We use the direct double sum as an independent
numerical check.

\subsubsection{Why a Fixed Residual Ceiling Fails}
For a stored template $\hat{\bm U}$, define the observed aligned
and residual energies as
\begin{equation}
	S=\norm{\hat{\bm U}^H\bm y}^2,\qquad R=\norm{\hat{\bm Q}\bm y}^2.\label{eq:SR}
\end{equation}
A fixed-residual test requires $R\le h$ for a finite threshold $h$.
With an exact reference this can be a suitable rule. With a noisy
reference, part of the legitimate signal falls into the residual
subspace and grows with the received energy.

\begin{proposition}[Residual-ceiling inconsistency]\label{pr:ceiling}
	Assume $m_j\ge2$ and $\tau_j<\infty$ for every region. Scale the
	legitimate gain vector as $\bm\beta=\sqrt E\bm b$, where the
	direction is fixed and $\norm{\bm b}=1$. Every test that requires
	$R\le h<\infty$ has $P_{\FR}\to1$ as $E\to\infty$.
\end{proposition}
\begin{proof}
	In each region, $C_j<1$ almost surely. The regions have disjoint
	supports, so $\norm{\hat{\bm Q}\bm U\bm b}^2=
	\sum_j|b_j|^2(1-C_j)>0$ almost surely. Given the template,
	$R/E$ converges in probability to this positive value because
	the noise divided by $\sqrt E$ vanishes. Thus
	$\Prob(R\le h\mid\hat{\bm U})\to0$. Bounded convergence gives
	the population result.
\end{proof}

We use an \emph{energy-gated angular rule} that scales the allowed
residual energy with the aligned energy:
\begin{equation}
	\text{accept}\quad\Longleftrightarrow\quad S\ge g\ \text{and}\ R\le\xi S,
	\qquad \eta=(1+\xi)^{-1}.\label{eq:test}
\end{equation}
The angular condition can also be written as $S/(S+R)\ge\eta$.
It tests the fraction of observed energy in the enrolled subspace.
The aligned-energy gate also requires sufficient energy in
that subspace. This reduces acceptance of weak, noise-only
observations that satisfy the angular condition. The analysis
calibrates the combined test without asserting that either
gate is optimal on its own.

\subsubsection{Exact Conditional and Population Probabilities}
For a fixed mean and template, the aligned and residual energies
of the mean are
\begin{equation}
	s=\norm{\hat{\bm U}^H\bm\mu}^2,\quad t=\norm{\hat{\bm Q}\bm\mu}^2.
	\label{eq:st}
\end{equation}
The projected noise components are independent because
$\hat{\bm P}$ and $\hat{\bm Q}$ project onto orthogonal subspaces
and the receiver noise has identity covariance. It follows that
\begin{equation}
	2S\sim\chi'^2_{2q}(2s),\qquad 2R\sim\chi'^2_{2k}(2t).\label{eq:conditionalchi}
\end{equation}
Let $f_{S;s}$ be the density of $S$. Conditioning on $S$ gives the
exact acceptance probability
\begin{equation}
	\C(s,t)=\int_g^\infty f_{S;s}(x)F_{\chi'^2_{2k}(2t)}(2\xi x)\,dx.\label{eq:accept}
\end{equation}
Increasing $s$ makes the two conditions in \eqref{eq:test} easier
to meet, while increasing $t$ makes the residual condition harder.
These statements follow from stochastic monotonicity of the
noncentral chi-square distributions and the shape of the
acceptance region. For a fixed mean, $S$ and $R$ are independent conditional
on the template. Averaging over enrollment generally introduces
dependence between them.

For a fixed mean, write its component on region $j$ as
$\bm\mu_j=\sqrt{e_j}\bm v_j$, with zeros outside that region.
Here $e_j=\norm{\bm\mu_j}^2$. For $e_j>0$, $\bm v_j$ is a unit
vector and $c_j=|\bm u_j^H\bm v_j|^2$. Proposition~\ref{pr:overlap}
gives the law of $Z_j=|\hat{\bm u}_j^H\bm v_j|^2$. Regions with
$e_j=0$ contribute zero to both sums below. The overlaps are independent across
regions, so averaging \eqref{eq:accept} yields
\begin{equation}
	\Prob(\text{accept})=\E_{\bm Z}\!\left[
	\C\!\left(\sum_j e_j Z_j,\sum_j e_j(1-Z_j)\right)\right].\label{eq:mixture}
\end{equation}
For Alice, $Z_j=C_j$ and $e_j=E_j$. Equation~\eqref{eq:mixture}
is exact when the mean is fixed independently of enrollment.
An optimized beamformer that uses the realized template can
change the direction and hence the overlap laws. We condition
on the template for that case in Section~\ref{sec:adaptive}.

\begin{corollary}[Enrollment-limited verification]\label{cor:floor}
	For $q=1$, a fixed direction with overlap $Z$ and received
	energy $E$ has limiting acceptance probability
	\begin{equation}
		\lim_{E\to\infty}\Prob(\text{accept})=1-F_Z(\eta).
		\label{eq:floor}
	\end{equation}
	Alice's limiting FR is $F_C(\eta)$ and can remain positive.
	At $\tau=0$, $Z\sim\operatorname{Beta}(1,M-1)$ for every $c$.
	In that case, the test cannot distinguish fixed directions
	with equal energy. As $\tau\to\infty$, $Z\to c$ in probability.
\end{corollary}
\begin{proof}
	Given the template, $S/E\to Z$, $R/E\to1-Z$, and $g/E\to0$.
	The overlap has no probability mass at $\eta$. Bounded convergence
	therefore gives~\eqref{eq:floor}. The claim at $\tau=0$ follows
	from the isotropy of enrollment noise. The large-$\tau$ claim
	follows from consistency of~\eqref{eq:enrollment}.
\end{proof}
With several regions, let $\omega_j=e_j/\sum_i e_i$ be the fixed
fractions of a nonzero mean's energy, so $\omega_j\ge0$ and $\sum_j\omega_j=1$. The limit
becomes $\Prob(\sum_j\omega_j Z_j>\eta)$. This enrollment floor
applies to the specified verifier and observation model.

\begin{figure}[t]
	\centering
	\subfloat[]{%
		\includegraphics[width=0.47\columnwidth]
		{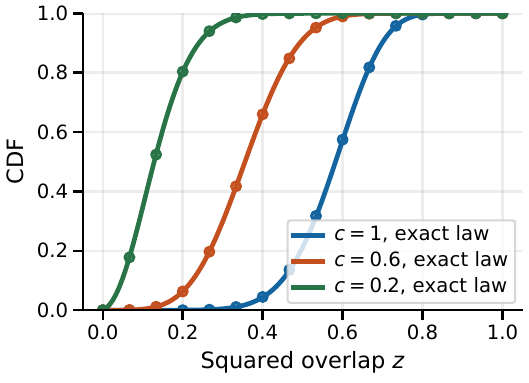}
		\label{fig:enrollment:a}%
	}
	\hfill
	\subfloat[]{%
		\includegraphics[width=0.47\columnwidth]
		{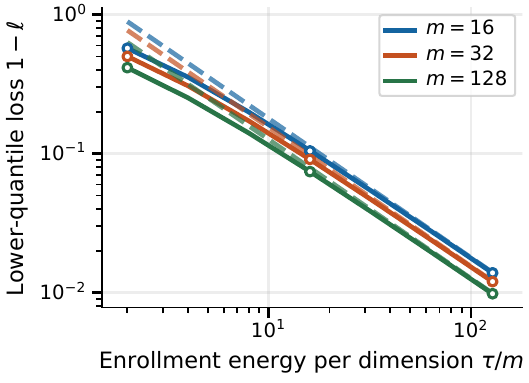}
		\label{fig:enrollment:b}%
	}
	\caption{Enrollment overlap and its lower-tail approximation.
		(a) Overlap CDFs for $m=16$ and $\tau=20$. Blue, orange, and
		green denote $c=1$, $0.6$, and $0.2$, respectively. Solid
		curves show \eqref{eq:betacdf}. Open circles are empirical
		CDF values from $10^5$ independent complex enrollment vectors
		per curve.
		(b) Lower-tail overlap loss for $\delta_j=0.002$. Blue, orange,
		and green denote $m=16$, $32$, and $128$, respectively.
		Solid curves give the exact quantiles in \eqref{eq:ell} and
		dashed curves give the approximation in \eqref{eq:quantileasym}.
		Open circles at $\tau/m=2,16,128$ show sample quantiles from
		$10^5$ independent enrollments per point. Error bars give
		pointwise $95\%$ intervals from the binomial order statistics.}
	\label{fig:enrollment}
\end{figure}

\subsection{FR Control Uniform in Legitimate Gain}
\subsubsection{A High-Probability Enrollment Event}
We write $F_{C;\tau}$ for the CDF of the legitimate overlap $C$
at received enrollment energy $\tau$. For region $j$,
$F_{C_j;\underline{\tau}_j}$ is the CDF of $C_j$ evaluated
under the enrollment energy $\underline{\tau}_j$.
Choose an enrollment-outage budget $\delta_j\in(0,1)$ for each
region. The corresponding lower overlap quantile is
\begin{equation}
	\ell_j=F^{-1}_{C_j;\lo\tau_j}(\delta_j).
	\label{eq:ell}
\end{equation}
The lower bound common to all regions is
\begin{equation}
	\ell=\min_j\ell_j,
	\label{eq:minimum_overlap}
\end{equation}
and the overall enrollment-outage budget is
\begin{equation}
	\delta=1-\prod_j(1-\delta_j).
	\label{eq:overall_outage}
\end{equation}
The overlap $C_j$ is stochastically increasing in $\tau_j$.
The quantiles are therefore conservative for $\tau_j\ge\lo\tau_j$.
Independence across regions gives
\begin{equation}
	\mathcal G=\{C_j\ge\ell_j,\ \forall j\},\quad \Prob(\mathcal G)\ge1-\delta.
	\label{eq:good}
\end{equation}
For any template in $\mathcal G$, every legitimate gain vector
gives the energy bounds
\begin{equation}
	s=\sum_jE_j C_j\ge\ell E,\quad
	t=\sum_jE_j(1-C_j)\le(1-\ell)E.\label{eq:goodenergies}
\end{equation}
The event $\mathcal G$ serves as a certificate in the analysis.
Bob can set the thresholds without observing the exact $C_j$.

\subsubsection{Constructive Thresholds}
Choose $\alpha_g,\alpha_r>0$ such that $\alpha_g+\alpha_r<1$.
Set the aligned-energy threshold to
\begin{equation}
	g=\tfrac12 F^{-1}_{\chi'^2_{2q}(2\ell E_{\min})}(\alpha_g).
	\label{eq:g}
\end{equation}
For an angular factor $\xi>(1-\ell)/\ell$, the residual-gate
bound is defined by
\begin{align}
	v_{\max}&=\ell-(1-\ell)/\xi,\label{eq:vmax}\\
	\mathcal B(\xi)&=\inf_{0<v<v_{\max}}\exp\!\left\{
	-k\log(1-v)-q\log(1+\xi v)\right.\nonumber\\[-1mm]
	&\hspace{14mm}\left.+E_{\min}\!\left[\frac{(1-\ell)v}{1-v}
	-\frac{\ell\xi v}{1+\xi v}\right]\right\}.\label{eq:frchernoff}
\end{align}
Choose $\xi$ so that $\mathcal B(\xi)\le\alpha_r$.
The inner minimization is convex in the single variable $v$.
A bracketed search over $\xi$ sets the angular threshold using
the bound, without training data or attacker samples.

\begin{theorem}[Gain-uniform calibration]\label{th:fr}
	Use the calibration in \eqref{eq:ell}--\eqref{eq:frchernoff}.
	For every realized template in $\mathcal G$,
	\begin{equation}
		\sup_{\bm\beta:\norm{\bm\beta}^2\ge E_{\min}}
		P_{\FR}(\bm\beta\mid\hat{\bm U})\le\alpha_g+\alpha_r.
		\label{eq:conditionalfr}
	\end{equation}
	Averaging over enrollment gives the following bound even when
	the legitimate gains depend on the realized template:
	\begin{equation}
		P_{\FR}\le\delta+(1-\delta)(\alpha_g+\alpha_r).
		\label{eq:unconditionalfr}
	\end{equation}
	The bounds hold for arbitrarily large legitimate verification energy.
\end{theorem}
\begin{proof}
	Fix a template in $\mathcal G$ and a legitimate gain vector.
	By stochastic monotonicity, the bound on $s$ in
	\eqref{eq:goodenergies} and the threshold in \eqref{eq:g} give
	$\Prob(S<g)\le\alpha_g$. For $0<v<1$, conditional independence in \eqref{eq:conditionalchi}
	gives the moment-generating function (MGF) bound
	\begin{align}
		\Prob(R>\xi S)&\le\E [e^{v(R-\xi S)}]\nonumber\\
		&=\frac{\exp\{tv/(1-v)-s\xi v/(1+\xi v)\}}
		{(1-v)^k(1+\xi v)^q}.\label{eq:mgffr}
	\end{align}
	At a fixed $E=s+t$, increasing $s$ decreases the exponent.
	Substituting the lower bound $s\ge\ell E$ gives a coefficient
	of $E$ that satisfies
	\begin{equation}
		\frac{(1-\ell)v}{1-v}-\frac{\ell\xi v}{1+\xi v}\le0
		\quad\Longleftrightarrow\quad v\le v_{\max}.\label{eq:negativecoefficient}
	\end{equation}
	Since this coefficient is nonpositive, using $E_{\min}$ in place
	of $E$ gives an upper bound in \eqref{eq:mgffr} for every
	$E\ge E_{\min}$. A union bound over the two ways to reject gives
	\eqref{eq:conditionalfr}. On $\mathcal G^c$ we bound the
	rejection probability by one. Averaging gives
	\eqref{eq:unconditionalfr}.
\end{proof}

This calibration gives sufficient thresholds. Exact minimax
thresholds would require a separate optimization. The analysis
uses simultaneous regional quantiles, replaces the regional
overlaps by their minimum, and applies Chernoff and union bounds.
These steps can make the calibration conservative.
To see why a feasible $\xi$ exists for any positive $\alpha_r$,
take $v=\xi^{-1/2}$ at sufficiently large $\xi$ in
\eqref{eq:frchernoff}. The factor $(1+\xi v)^{-q}$ vanishes
while the rest of the exponential stays bounded. A feasible
calibration can still offer weak discrimination if enrollment
or verification energy is too low.

\begin{figure}[t]
	\centering
	\subfloat[]{%
		\includegraphics[width=0.47\columnwidth]
		{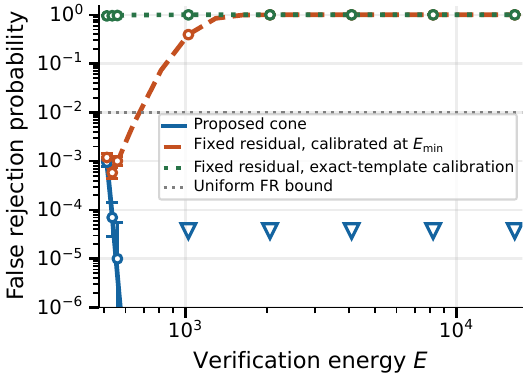}
		\label{fig:gain:a}%
	}
	\hfill
	\subfloat[]{%
		\includegraphics[width=0.47\columnwidth]
		{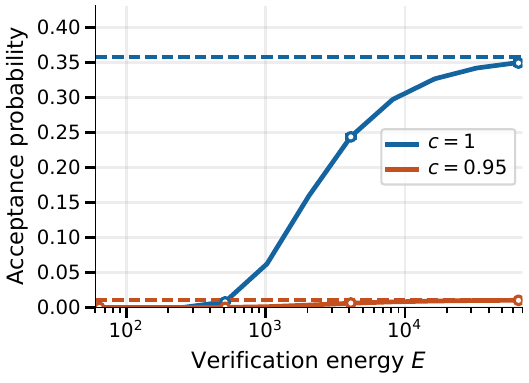}
		\label{fig:gain:b}%
	}
	\caption{Effect of verification energy with noisy enrollment.
		(a) FR versus verification energy for $M=128$, $\tau=1024$,
		and $E_{\min}=512$. The blue solid curve gives the cone rule
		from \eqref{eq:accept} and \eqref{eq:mixture}. The orange
		dashed curve uses the fixed ceiling in \eqref{eq:fixedbaseline}.
		The green dotted curve uses a fixed ceiling calibrated with an
		exact template. The horizontal gray dotted line gives the
		uniform FR bound in \eqref{eq:unconditionalfr}.
		(b) Acceptance with $m=32$, $\tau=256$, $g=10$, and $\eta=0.9$.
		Blue and orange denote $c=1$ and $c=0.95$, respectively.
		Solid curves give the enrollment averages in \eqref{eq:mixture}.
		Horizontal dashed lines give the limits in \eqref{eq:floor}.
		Open circles and error bars show full-vector Monte Carlo
		estimates and pointwise $95\%$ Clopper--Pearson intervals
		for all three receivers in (a) and both overlaps in (b).
		In panel~(a), open downward triangles
		mark the upper endpoints of these intervals at zero counts.
		A fixed ceiling can reject Alice at high verification energy.
		The angular test still shows the effect of enrollment noise
		at high verification energy.}
	\label{fig:gain}
\end{figure}

\subsection{A Template-Aware Adversary}\label{sec:adaptive}
\subsubsection{Exact Attainable-Energy Characterization}
For a fixed enrolled subspace, define the two channel matrices
\begin{equation}
	\bm A=\bm H_E^H\hat{\bm P}\bm H_E,\qquad
	\bm B=\bm H_E^H\hat{\bm Q}\bm H_E.\label{eq:AB}
\end{equation}
A precoder $\bm w$ gives $s=\bm w^H\bm A\bm w$ and
$t=\bm w^H\bm B\bm w$. To find the smallest residual energy
for a given $s$, use a Hermitian matrix
$\bm X\in\mathbb C^{L\times L}$ and define
\begin{align}
	t_P(s)=\min_{\bm X\succeq0}\ &\tr(\bm B\bm X)\nonumber\\
	\text{s.t. }&\tr(\bm A\bm X)=s, \nonumber\\ &\tr\bm X\le P,
	\label{eq:frontier}
\end{align}
for $0\le s\le s_{\max}=P\lambda_{\max}(\bm A)$.

\begin{theorem}[Physically attainable optimum]\label{th:frontier}
	For $s>0$, the problem in \eqref{eq:frontier} has a rank-one
	solution. Its value is the smallest residual energy that
	a deterministic beamformer can attain with budget $P$. The
	maximum FA probability is
	\begin{equation}
		P_{\FA}^*(\hat{\bm U})=
		\max_{0\le s\le s_{\max}}\C(s,t_P(s)).\label{eq:optimal}
	\end{equation}
	The frontier is convex and nondecreasing. On any interval
	$[a,b]$ in its domain, the objective obeys
	\begin{equation}
		\max_{s\in[a,b]}\C(s,t_P(s))\le\C(b,t_P(a)).\label{eq:intervalbound}
	\end{equation}
\end{theorem}
\begin{proof}
	The feasible set is compact, so there is an optimal extreme point.
	There are at most two active real affine constraints. The
	complex Hermitian rank bound for semidefinite programming
	(SDP) gives $r^2\le2$~\cite{huang2010}, so $r=1$ when $s>0$.
	For $s=0$, $\bm w=0$ attains $t_P(0)=0$. The matrix formulation
	therefore gives the same lower frontier as the deterministic
	precoders. Since acceptance decreases with residual energy,
	the frontier point at a given aligned energy is at least as
	good as any feasible beamformer at that energy. This proves
	\eqref{eq:optimal}. Convex combinations of feasible matrices
	give convexity of the frontier. Scaling a beamformer at
	$s_2>s_1$ gives $t_P(s_1)\le(s_1/s_2)t_P(s_2)\le t_P(s_2)$.
	The monotonicity of $\C$ gives~\eqref{eq:intervalbound}.
\end{proof}

The standard rank bound reduces the noisy-template probability
optimization to a search along the energy frontier. The search
evaluates a feasible probability at each interval midpoint and
uses \eqref{eq:intervalbound} to bound the probability over that
interval. It then refines the largest unresolved interval. This
gives a bracket on the global optimum. The implementation uses
quadrature and floating-point optimization. A formal certificate
for the reported numerical tolerance would require interval
arithmetic.

For $q=1$, set $\bm a=\bm H_E^H\hat{\bm u}$, so that
$\bm A=\bm a\bm a^H$. Away from the zero-residual null-space
case, the frontier beamformer takes the regularized form
\begin{equation}
	\bm w_\nu=\frac{\sqrt{s}(\bm B+\nu\bm I)^{-1}\bm a}
	{\bm a^H(\bm B+\nu\bm I)^{-1}\bm a},\quad \nu>0.\label{eq:secular}
\end{equation}
Here, $\nu$ is the Lagrange multiplier associated with the
transmit-energy constraint, and it acts as a regularization
parameter.
For $s>0$ and $\bm a\ne\bm0$, use the minimum-energy zero-residual
solution if it meets the power constraint. Otherwise, the
$\nu=0$ pseudoinverse solution applies only if
$\bm a\in\range(\bm B)$ and the resulting precoder meets the
constraint. In the remaining interior cases, find $\nu>0$
from $\norm{\bm w_\nu}^2=P$. At $s=s_{\max}>0$, the solution
is $\sqrt P\bm a/\norm{\bm a}$. If $\bm a=\bm0$, the only
attainable aligned energy is zero and the silent precoder
attains the lower frontier. Independent SDP solutions can
be used to check these cases. With one transmit antenna and
any $q$, the frontier is $t_P(s)=Bs/A$ when $A>0$.

\subsubsection{A Convex Upper Certificate for Any Verifier Rank}
Repeated frontier searches can be expensive across a range of
designs. The following upper bound can be evaluated separately.
For $\lambda_0,v\ge0$, set $a=\lambda_0+\xi v<1$ and define
\begin{align}
	\bm D(a,v)&=\frac{a}{1-a}\bm A-\frac{v}{1+v}\bm B,\label{eq:D}\\
	\mathcal L(\lambda_0,v)&=-\lambda_0g-q\log(1-a)-k\log(1+v)\nonumber\\
	&\quad+P\max\{0,\lambda_{\max}(\bm D(a,v))\}.\label{eq:logbound}
\end{align}

\begin{theorem}[Convex adversarial Chernoff certificate]\label{th:chernoff}
	For an attacker that knows the realized template,
	\begin{equation}
		P_{\FA}^*(\hat{\bm U})\le
		\exp\!\left\{\inf_{\lambda_0,v\ge0:\lambda_0+\xi v<1}
		\mathcal L(\lambda_0,v)\right\}.\label{eq:certificate}
	\end{equation}
	The objective is convex. Every feasible pair gives a valid
	upper bound, and minimizing the objective tightens the bound.
\end{theorem}
\begin{proof}
	On acceptance, $\lambda_0(S-g)+v(\xi S-R)\ge0$.
	Applying the MGF bound to this event gives
	\begin{align}
		\Prob(\text{accept})
		&\le\E [e^{\lambda_0(S-g)+v(\xi S-R)}]\nonumber\\
		&=\frac{e^{-\lambda_0g}}{(1-a)^q(1+v)^k}
		\exp\!\left\{\frac{as}{1-a}-\frac{vt}{1+v}\right\}.\label{eq:attackmgf}
	\end{align}
	The maximum of the exponent over $\norm{\bm w}^2\le P$ is
	$P\max\{0,\lambda_{\max}(\bm D)\}$, where the maximization
	includes the silent attack $\bm w=0$. Each feasible pair of
	multipliers bounds every feasible beamformer. Taking the
	infimum therefore preserves the bound without exchanging
	the order of the two optimizations.
	
	To show convexity, write
	\begin{equation}
		\bm D=\frac{\bm A}{1-a}+\frac{\bm B}{1+v}-(\bm A+\bm B).\label{eq:matrixconvex}
	\end{equation}
	Each reciprocal is convex on its domain and is weighted by a
	positive semidefinite matrix. The resulting expression is
	matrix convex. Taking the largest eigenvalue and then its
	maximum with zero preserves convexity. The other terms in
	\eqref{eq:logbound} are convex as well.
\end{proof}

To solve the problem with a matrix inequality, replace the last
term by $Pz$ with $z\ge0$ and impose
\begin{equation}
	\begin{bmatrix}
		z\bm I+\bm A+\bm B&\bm A^{1/2}&\bm B^{1/2}\\
		\bm A^{1/2}&(1-a)\bm I&\bm0\\
		\bm B^{1/2}&\bm0&(1+v)\bm I
	\end{bmatrix}\succeq0.\label{eq:schur}
\end{equation}
Since $1-a>0$ and $1+v>0$, the two lower diagonal blocks are
positive definite. Their Schur complement gives $z\bm I\succeq\bm D$
by \eqref{eq:matrixconvex}.
The origin gives a probability bound of one, so the optimized
bound is at most one. The certificate can still be loose.

\begin{figure}[t]
	\centering
	\subfloat[]{%
		\includegraphics[width=0.47\columnwidth]
		{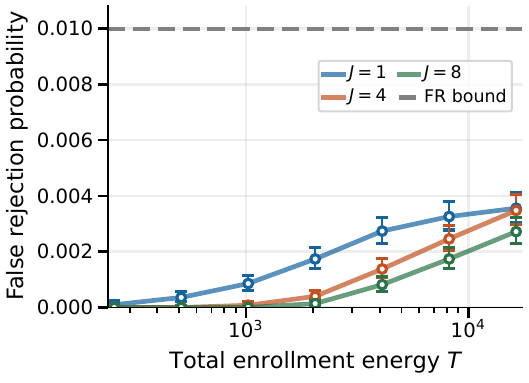}
		\label{fig:reliability-spoofing:a}%
	}
	\hfill
	\subfloat[]{%
		\includegraphics[width=0.47\columnwidth]
		{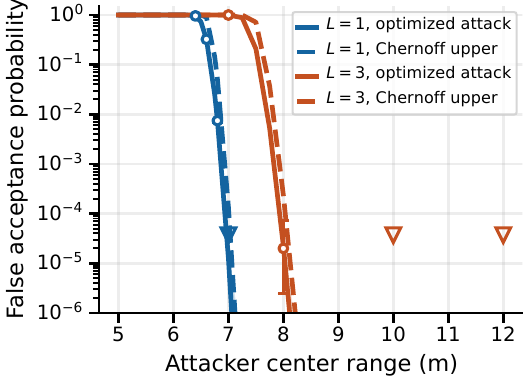}
		\label{fig:reliability-spoofing:b}%
	}
	\caption{Legitimate-user reliability and conditional spoofing performance.
		(a) Population FR versus total enrollment energy $T$ at
		$E=E_{\min}=512$. The enrollment and verification energies
		in each region are $T/J$ and $E/J$, respectively. Blue,
		orange, and green denote $J=1$, $4$, and $8$, respectively.
		Open circles and error bars show Monte Carlo estimates
		and pointwise $95\%$ Clopper--Pearson intervals from
		$5\times10^4$ independent enrollment and verification pairs
		per point. The colored lines connect these estimates. The
		gray dashed line gives the unconditional FR bound
		$\FRBound$ in \eqref{eq:unconditionalfr}.
		(b) FA versus attacker center range at $P=2048$, conditional
		on one fixed noisy template known to the attacker. Blue
		and orange denote $L=1$ and $L=3$, respectively. Solid
		curves give achievable probabilities from \eqref{eq:optimal}
		evaluated with \eqref{eq:accept}. Dashed curves of the
		same colors give the upper bounds in \eqref{eq:certificate}.
		Open circles and error bars show independent online-noise
		Monte Carlo estimates for selected precoders and pointwise
		$95\%$ Clopper--Pearson intervals. Open downward triangles
		mark the upper endpoints of these intervals at zero counts.
		Shading shows the numerical optimization brackets from
		\eqref{eq:intervalbound} with the stated numerical
		allowance. For $L=3$, the transmit antennas are offset
		laterally by $-0.05$, $0$, and $0.05$~m from the attacker
		center.}
	\label{fig:reliability-spoofing}
\end{figure}

\subsection{From Geometry to a Pre-Enrollment Certificate}
\subsubsection{Controlling a Template-Adaptive Attack Before Enrollment}
Let $\bm P_0=\bm U\bm U^H$ be the projector onto the true
legitimate subspace. Define the coherence of the attack
subspace with this subspace as
\begin{equation}
	c_0=\sup_{\bm\mu\in\range(\bm H_E),\,\norm{\bm\mu}=1}
	\norm{\bm U^H\bm\mu}^2.\label{eq:c0}
\end{equation}
Let the columns of $\bm V_E$ form an orthonormal basis for
$\range(\bm H_E)$. Then $c_0=\sigma_{\max}^2(\bm U^H\bm V_E)$,
where $\sigma_{\max}$ is the largest singular value. For a
numerically rank-deficient channel, the computed rank depends
on the tolerance used for the singular values.

\begin{lemma}[Enrollment-induced angle inflation]\label{lem:angle}
	On $\mathcal G$, every unit vector $\bm\mu$ in the attack
	subspace satisfies $\norm{\hat{\bm U}^H\bm\mu}^2\le\hi c$, where
	\begin{equation}
		\hi c=\begin{cases}
			\left(\sqrt{c_0\ell}+\sqrt{(1-c_0)(1-\ell)}\right)^2,&c_0<\ell,\\
			1,&c_0\ge\ell.
		\end{cases}\label{eq:inflation}
	\end{equation}
\end{lemma}
\begin{proof}
	Since the regions have disjoint supports, $\bm U^H\hat{\bm U}$
	is diagonal and the squared magnitudes of its diagonal entries
	are $C_j$. On $\mathcal G$, the largest principal angle between
	the true and enrolled subspaces is at most
	$\theta_\ell=\arccos\sqrt\ell$. Every unit attack vector makes
	an angle of at least $\theta_{sep}=\arccos\sqrt{c_0}$ with the true
	legitimate subspace. Its angle to the enrolled subspace is
	therefore at least $(\theta_{sep}-\theta_\ell)_+$. Expanding the
	squared cosine gives \eqref{eq:inflation}. The angle inequality
	also follows by resolving the vector along its closest
	subspace direction and in the orthogonal complement, then
	applying Cauchy--Schwarz.
\end{proof}

The maximum received attack energy is
$E_{\max}^E=P\lambda_{\max}(\bm H_E^H\bm H_E)$. On $\mathcal G$,
every template-aware attack satisfies $s\le\hi c(s+t)$ and
$s+t\le E_{\max}^E$. These inequalities give the scalar objective
\begin{align}
	\mathcal L_0(\lambda_0,v)&=-\lambda_0g-q\log(1-a)-k\log(1+v)\nonumber\\
	&\quad+E_{\max}^E\max\left\{0,\frac{\hi c a}{1-a}
	-\frac{(1-\hi c)v}{1+v}\right\}.\label{eq:envelope}
\end{align}
The reciprocal representation used in \eqref{eq:matrixconvex}
also proves convexity here. Set $B_0=\exp(\inf\mathcal L_0)$,
taking the infimum over the domain in Theorem~\ref{th:chernoff}.

\begin{corollary}[Unconditional adaptive-attack certificate]\label{cor:unconditional}
	Fix the physical channel $\bm H_E$. For every measurable
	attack policy $\bm w(\hat{\bm U})$ with budget $P$,
	\begin{equation}
		\E_{\hat{\bm U}}[P_{\FA}(\bm w(\hat{\bm U})\mid\hat{\bm U})]
		\le\delta+(1-\delta)B_0.\label{eq:outagecert}
	\end{equation}
\end{corollary}
\begin{proof}
	At fixed received energy $e=s+t>0$, the exponent in
	\eqref{eq:attackmgf} increases with $s/e$. On $\mathcal G$,
	we upper-bound this ratio by $\hi c$. Maximizing over
	$0\le e\le E_{\max}^E$ includes the zero-energy case and
	gives \eqref{eq:envelope}. The bound holds for every feasible
	attack and every template in $\mathcal G$. We average over
	enrollment and use the bound of one on $\mathcal G^c$.
\end{proof}
The scalar bound treats maximum received energy and maximum
coherence separately. It can therefore be looser than the
conditional matrix certificate. It bounds each adaptive
attack on $\mathcal G$ before averaging. The beta mixture
in \eqref{eq:mixture} applies to attacks fixed independently
of enrollment. If the attacker can move over a spatial set,
the guarantee needs a channel or subspace bound that holds
throughout that set. A range sweep provides values at the
sampled positions only.

\subsubsection{The Enrollment Cost of Angular Separation}
The angle bound gives the enrollment energy needed to preserve
angular separation. Consider one legitimate region of dimension
$m$ with a fixed cone threshold $\eta$. Assume $c_0<\eta$.
The available angular margin is
\begin{equation}
	\psi=\arccos\sqrt{c_0}-\arccos\sqrt\eta>0.
	\label{eq:margin}
\end{equation}
The coherence bound on $\mathcal G$ stays at or below $\eta$
if $\ell\ge\cos^2\psi$. Define the exact sufficient
enrollment budget as
\begin{equation}
	\tau_\delta(\psi)=\inf\{\tau\ge0:
	F_{C;\tau}(\cos^2\psi)\le\delta\}.\label{eq:trainingbudget}
\end{equation}
We compute this budget by inverting the known overlap CDF
as a monotone function of $\tau$.

\begin{theorem}[Inverse-square certification budget]\label{th:cost}
	Fix $m\ge2$ and $\delta\in(0,1)$. Let $G_{m-1}^{-1}(1-\delta)$
	be the $(1-\delta)$ quantile of a Gamma distribution with
	shape $m-1$ and unit scale. Then
	\begin{equation}
		\lim_{\psi\downarrow0}\psi^2\tau_\delta(\psi)=G_{m-1}^{-1}(1-\delta).
		\label{eq:costlaw}
	\end{equation}
	Equivalently, the lower overlap quantile at large $\tau$ satisfies
	\begin{equation}
		1-\ell=\frac{G_{m-1}^{-1}(1-\delta)}{\tau}+o(\tau^{-1}).\label{eq:quantileasym}
	\end{equation}
\end{theorem}
\begin{proof}
	Use coordinates with the first axis along $\bm u$. Set
	$X=|\sqrt\tau+\epsilon_1|^2$ and
	$Y=\sum_{i=2}^m|\epsilon_i|^2$. Then
	$Y\sim\operatorname{Gamma}(m-1,1)$ and $X/\tau\to1$ in probability.
	Consequently,
	\begin{equation}
		\tau(1-C)=\frac{Y}{X/\tau+Y/\tau}\ \Longrightarrow\ Y.\label{eq:gamlimit}
	\end{equation}
	The Gamma CDF is continuous and strictly increasing on its
	support. Its fixed interior quantiles are thus the limits
	of those of the scaled overlap loss, which gives
	\eqref{eq:quantileasym}. Substituting
	$1-\cos^2\psi=\sin^2\psi\sim\psi^2$ into the quantile condition
	gives \eqref{eq:costlaw}.
\end{proof}
This budget is sufficient for the angular envelope at the
stated threshold. A protocol-wide necessary energy or a
finite-energy FA target would require a separate analysis.
The energy gate and receiver noise still enter through
\eqref{eq:envelope}. If the receiver recalibrates $\eta$ as
$\tau$ changes, solve for the two together and recheck the margin
before applying the fixed-threshold approximation. For $J>1$,
use the regional budgets $\delta_j$ and retain the smallest
overlap bound.

\subsection{Implementation and Computational Cost}
The receiver uses the enrolled template to authenticate a packet.
The attacker channel is used only in the offline evaluation of
the threat model. The receiver operates in three steps.
\begin{enumerate}
	\item \emph{Design.} Fix the regional partition and the trusted
	lower energy bounds. Choose $(\delta_j,\alpha_g,\alpha_r)$.
	Compute the noncentral-$F$ overlap quantiles in
	\eqref{eq:ell} and obtain $(g,\xi)$ from
	\eqref{eq:g}--\eqref{eq:frchernoff}. Store both thresholds.
	\item \emph{Trusted enrollment.} Acquire one noise-affected
	observation per region by combining the trusted pilots.
	Normalize it using \eqref{eq:enrollment_direction}.
	The array partition fixes region membership even when a
	noiseless entry is zero. Every observed element contributes
	enrollment noise.
	\item \emph{Online decision.} Compute the $J$ regional inner
	products and sum their squared magnitudes to get $S$.
	Compute $R=\norm{\bm y}^2-S$ and apply \eqref{eq:test}.
	Reuse the stored template until the protocol calls for a
	new, independently trusted enrollment.
\end{enumerate}
The disjoint supports let the receiver use regional inner
products in place of a dense $M\times M$ projector. The
decision, template storage, and enrollment normalization
each scale as $O(M)$. Offline calibration uses scalar quantile
inversion and scalar convex minimization. Its numerical cost
depends on the noncentralities and the requested tolerance.

Forming $\bm A,\bm B$ with direct Gram products costs
$O(ML^2)$ offline. Evaluating \eqref{eq:logbound} requires the
largest eigenvalue of an $L\times L$ Hermitian matrix, which
costs $O(L^3)$ with a dense implementation. The alternative
formulation uses a $3L\times3L$ matrix inequality. The exact
frontier search also evaluates \eqref{eq:accept} and refines
a one-dimensional interval list. The number of iterations
depends on the data and the tolerance. These costs enter the
offline analysis and leave the online decision linear in $M$.

The scalar certificate applies before an actual template is
available. Once enrollment is complete, the matrix certificate
can use the realized geometry and is often tighter. Both
computations assess the same fixed decision rule without fitting
it to simulated attacks.

\section{Simulation}
We use spherical-wave channels to check the probability laws and
compare the receiver under different enrollment, gain, and attack
conditions.
The simulation codes are available in~\cite{rajabi2026enrollmentcode}.

\subsection{Configuration and Evidence Types}
We use $M=128$ elements at a carrier frequency of $28$~GHz
with half-wavelength spacing. Alice is at broadside and
$R_0=r_A=5$~m, where $r_A$ denotes her range. We set $b_m=1$.
The aperture is about $0.680$~m and the Rayleigh distance
is about $86.3$~m. All transmitter ranges used here
lie in the radiative near field. These are idealized
line-of-sight simulations. Array coupling, calibration errors
outside the regional-gain model, and multipath could change
performance on measured channels.

Unless stated otherwise, we use $E_{\min}=512$, a total
lower enrollment energy of $T=4096$, $\lo\tau_j=T/J$,
and $P=2048$. We divide the enrollment-outage budget
$\delta=0.002$ equally across regions in the product in
\eqref{eq:overall_outage}, so $\delta_j=1-(1-\delta)^{1/J}$.
We set $\alpha_g=\alpha_r=0.004$. For $J=1$, the resulting
overlap bound and thresholds are
\begin{equation}
	\ell=\CalibrationEll,\quad g=\CalibrationG,\quad
	\xi=\CalibrationXi,\quad\eta=\CalibrationEta.
	\label{eq:numericalthresholds}
\end{equation}
The unconditional FR bound is $\FRBound$. Unless stated otherwise, we set both thresholds from the analytical
bounds before testing.

The original simulation run used \MCTotal\ online observations
and \MCEnroll\ enrollment vectors. Population trials draw fresh
enrollment and verification noise. Conditional trials fix the
template and precoder and draw new verification noise.
In the original run, the phase and threshold sweeps use $5\times10^4$
trials per point. Elsewhere we use $10^5$ trials unless noted.
Error bars give pointwise two-sided $95\%$ Clopper--Pearson
intervals. On logarithmic axes, a triangle marks the upper limit
at zero errors. The empirical CDF in Fig.~\ref{fig:enrollment}(a)
also admits a uniform Dvoretzky--Kiefer--Wolfowitz band.

We integrate \eqref{eq:accept} in standardized coordinates.
The absolute quadrature tolerance is $2\times10^{-9}$, and
the omitted $S$ tail has mass at most $2\times10^{-12}$.
The nominal allowance for the truncated Poisson CDF is
$2\times10^{-13}$. The attack search first targets a probability
gap of $2\times10^{-4}$. When the lower value is below $0.01$,
we refine unresolved intervals to
$\max\{2\times10^{-7},0.03p_{\rm coarse}\}$, where $p_{\rm coarse}$
is the initial achievable probability. We also record a
numerical allowance of $2\times10^{-8}$. Rare-event
probabilities below this numerical resolution remain
unresolved. Chernoff bounds may lie below the resolution
of the Monte Carlo estimates and are reported as upper bounds.

\subsection{Enrollment Noise and Verification Power}
In Fig.~\ref{fig:enrollment}(a), the exact CDFs agree with
Monte Carlo samples of the full complex enrollment vector.
In panel~(b), the exact quantile lies inside the empirical
interval at all nine sampled settings. The first-order approximation
improves as enrollment energy grows. It overestimates the loss at low $\tau/m$.
We use the exact noncentral-$F$ quantile for calibration.

Figure~\ref{fig:gain}(a) holds $T=1024$ and varies the
verification energy from $512$ to $16384$. The fixed-residual
baseline that accounts for enrollment noise uses the same
aligned-energy gate as the cone rule. Its residual threshold is
\begin{equation}
	h=\tfrac12F^{-1}_{\chi'^2_{2k}(2(1-\ell)E_{\min})}(1-\alpha_r).
	\label{eq:fixedbaseline}
\end{equation}
The baseline is conservative at the design energy. As the
legitimate gain grows, its rejection probability tends to one.
The second baseline sets both thresholds for an exact reference
by using $\ell=1$ in \eqref{eq:g} and \eqref{eq:fixedbaseline}.
Its aligned-energy threshold is $431.13$, compared with $365.43$
for the other two rules. All three use the same noisy enrollment
model. We integrate the two rejection events directly and sample
the transition above $E=512$ on a finer grid. At $E=536$, the
cone rule rejected $7$ of $10^5$ trials. The calculated FR is
$7.31\times10^{-5}$.

Figure~\ref{fig:gain}(b) uses $m=32$, $\tau=256$, $g=10$,
and $\eta=0.9$. We fix the directions to examine how
acceptance changes with energy at a given overlap. The
limits from \eqref{eq:floor} are $\LegitFloor$ for $c=1$
and $\AttackFloor$ for $c=0.95$. The limits show why
raising verification energy alone cannot recover an
inaccurately enrolled direction.

Figure~\ref{fig:reliability-spoofing}(a) shows the population
FR for $J\in\{1,4,8\}$. We recompute the thresholds at
each $T$ using \eqref{eq:ell}, \eqref{eq:g}, and
\eqref{eq:frchernoff}. Better enrollment allows a tighter
cone. The observed FR stays below the uniform bound.
For $J=1$, $T=4096$, $E=512$, and a template with $C=\ell$,
\eqref{eq:accept} gives a conditional FR of $0.429\%$, compared
with the conditional bound of $0.8\%$ in \eqref{eq:conditionalfr}. FR can rise
toward its allowance as enrollment improves because the
recalibrated gates become stricter. A wider cone can
also admit more attacks.

\subsection{Informed Attacks and Regional Robustness}
Figure~\ref{fig:reliability-spoofing}(b) conditions on one
fixed noisy template known to the attacker. We use
\eqref{eq:optimal} for both the single-antenna and the
three-antenna cases. The Chernoff certificate bounds the
achievable curve from above and can reach one. In this
geometry, the three-antenna attacker maintains a higher
probability of passing over a wider range of positions.
It can reproduce more of the legitimate signature, even
at a different range. For this template, the three-antenna array weakens the range
discrimination provided by wavefront curvature. At $8$~m, two of
$10^5$ three-antenna trials passed. The pointwise $95\%$ FA
interval was $[2.42\times10^{-6},7.22\times10^{-5}]$.

For $q=1$, the MSE benchmark adapted from~\cite{hassani2026}
minimizes $\norm{\bm H_E\bm w-\sqrt{E_{\min}}\hat{\bm u}}^2$
subject to $\norm{\bm w}^2\le P$. It fits one target signature.
A smaller fitting error need not increase the probability of
passing both gates on the same channel and template.

In Fig.~\ref{fig:regional}(a), we apply the phases
$(-\phi,+\phi,-\phi,+\phi)$ to the four equal aperture quarters.
The $J=4$ model can represent these phases with one gain for
each quarter. The $J=1$ model assumes one gain for the entire
aperture, so the phase changes violate its shape assumption.
The four-region verifier retains a low FR in this experiment.
The one-region verifier rejects more often as the disturbance
grows.

Figure~\ref{fig:regional}(b) shows population FA for a
single broadside source at the same received signal energy as
Alice, $E=512$. At range $r$, the source transmits with energy
$512/\norm{\bm h(r,0)}^2$. This stays below $P=2048$ at all
eight ranges from $6$ to $9.5$~m. We recalibrate the two gates
for each $J$. More regional gains let the receiver accept larger
changes in the spatial signature. At $8$~m, it accepted $7490$
of $10^5$ trials for $J=4$ and $99212$ for $J=8$.

In Table~\ref{tab:paired}, both receivers use the same observations.
For each $J$, the cutoff for $S/(S+R)$ alone matches the two-gate
FR on $2\times10^5$ nominal calibration trials ($\phi=0$).
At $P=2048$, the received energies are $631.8$, $567.1$, $511.8$,
and $423.0$ in order of increasing range. Matching Alice's energy
at $10$ and $11$~m would require $2048.8$ and $2478.9$, both above
the transmit budget. At $11$~m, two-gate FA reached $14.75\%$
for $J=8$, while no trial passed for $J=4$. 
With both gates, $J=4$ offers the best trade-off among the tested
partitions, with lower FR than $J=1$ at $\phi=20^\circ$ and lower FA
than $J=8$ across the tested ranges.
At $J=4$, Angle trades slightly higher FR at $20^\circ$ for lower FA
at $9$~m.

Figure~\ref{fig:regional}(c) uses the same four-region trials.
The separate calibration tested $1812$ gate pairs on $2\times10^5$
trials per case. It minimized the larger FA across two reference
geometries, breaking ties by the sum, with a simultaneous one-sided
$95\%$ upper FR limit of $1\%$ at $20^\circ$ (Bonferroni-corrected
Clopper--Pearson). We then fixed both pairs. At $9$~m the selected
pair reduced FA from $1.084\%$ to $0.002\%$ and raised FR from
$0.1442\%$ to $0.7924\%$. At $11$~m, $71$ trials passed the
selected energy gate and none passed the angular gate.

\begin{table}[t]
	\centering
	\caption{Population FR and FA (\%) at $T=4096$, with $5\times10^5$
		trials per case. Both and Angle denote the two-gate and angular-only rules. The FA columns give broadside source ranges in meters.}
	\label{tab:paired}
	\footnotesize
	\setlength{\tabcolsep}{2.05pt}
	\begin{tabular}{@{}clrrrrrrrr@{}}
		\toprule
		& & \multicolumn{2}{c}{FR ($\phi$)} & \multicolumn{4}{c}{FA ($P=2048$)} & \multicolumn{2}{c}{FA ($E=512$)} \\
		\cmidrule(lr){3-4}\cmidrule(lr){5-8}\cmidrule(lr){9-10}
		$J$ & Rule & $0^\circ$ & $20^\circ$ & $9$ & $9.5$ & $10$ & $11$ & $9$ & $9.5$ \\
		\midrule
		$1$ & Both & 0.2672 & 74.58 & 0 & 0 & 0 & 0 & 0 & 0 \\
		$1$ & Angle & 0.2512 & 93.22 & 0 & 0 & 0 & 0 & 0 & 0 \\
		$4$ & Both & 0.1412 & 0.1442 & 1.084 & 0.029 & 0.0002 & 0 & 0.098 & 0.0086 \\
		$4$ & Angle & 0.1608 & 0.1544 & 0.0202 & 0 & 0 & 0 & 0.0012 & 0 \\
		$8$ & Both & 0.0766 & 0.0718 & 99.99 & 99.58 & 93.84 & 14.75 & 97.69 & 96.15 \\
		$8$ & Angle & 0.0778 & 0.075 & 98.32 & 86.26 & 56.89 & 8.2376 & 76.62 & 66.94 \\
		\bottomrule
	\end{tabular}
\end{table}

\begin{figure}[t]
	\centering
	\subfloat[]{%
		\includegraphics[width=0.47\columnwidth]
		{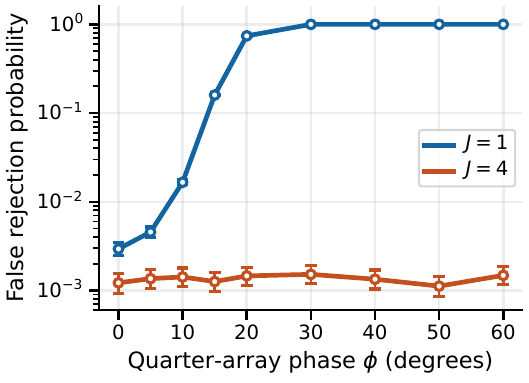}
		\label{fig:regional:a}%
	}
	\hfill
	\subfloat[]{%
		\includegraphics[width=0.47\columnwidth]
		{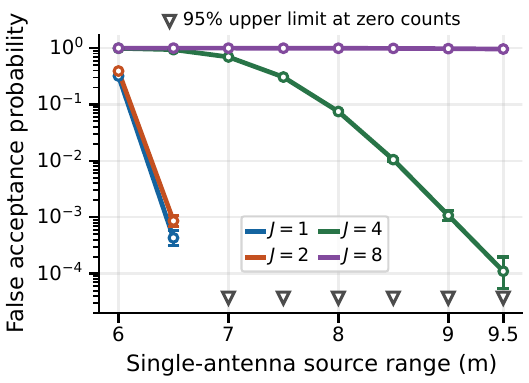}
		\label{fig:regional:b}%
	}
	\par
	\subfloat[]{%
		\begin{minipage}{0.98\columnwidth}
			\begin{minipage}[c]{0.585\linewidth}
				\includegraphics[width=\linewidth]{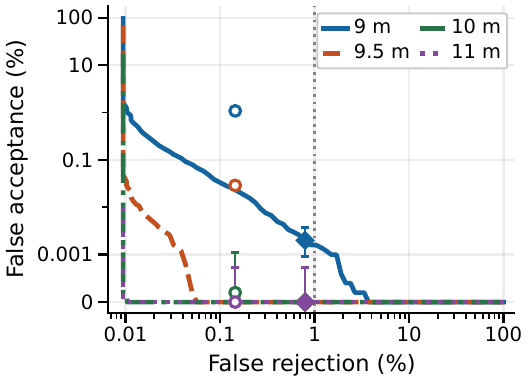}
			\end{minipage}\hfill
			\begin{minipage}[c]{0.405\linewidth}
				\centering\scriptsize
				\setlength{\tabcolsep}{1.45pt}
				\begin{tabular}{@{}lrr@{}}\toprule
					& Original & Selected \\ \midrule
					$g$ & 410.602 & 390.072 \\
					$\eta$ & 0.69637 & 0.73500 \\
					FR (\%) & 0.1442 & 0.7924 \\
					\midrule
					FA, $9$ m (\%) & 1.084 & 0.002 \\
					FA, $9.5$ m (\%) & 0.029 & 0 \\
					FA, $10$ m (\%) & 0.0002 & 0 \\
					FA, $11$ m (\%) & 0 & 0 \\
					\bottomrule
				\end{tabular}
			\end{minipage}
		\end{minipage}%
		\label{fig:regional:c}%
	}
	\caption{Regional gain variation and range discrimination.
		(a) FR under alternating quarter-array phases ($5\times10^4$ trials per point).
		(b) FA at matched received energy $512$ ($10^5$ trials per point).
		Lines join positive Monte Carlo estimates (circles). Gray triangles
		mark the common upper limit after zero acceptances for $J=1,2$
		at $7$--$9.5$~m.
		(c) $J=4$, $\phi=20^\circ$, and $P=2048$ for Eve.
		Each curve varies $\eta$ at the selected $g$ for one range.
		Open circles mark the original gates and diamonds the selected gates.
		The dotted vertical line marks $1\%$ FR. The curves and inset
		use $5\times10^5$ trials per case. Curves coincide where FA is zero.
		For zero acceptances in this panel or Table~\ref{tab:paired},
		the pointwise two-sided $95\%$ upper limit is $0.000738\%$.
		Error bars give pointwise $95\%$ Clopper--Pearson intervals.}
	\label{fig:regional}
\end{figure}

\begin{figure}[t]
	\centering
	\subfloat[]{%
		\includegraphics[width=0.47\columnwidth]
		{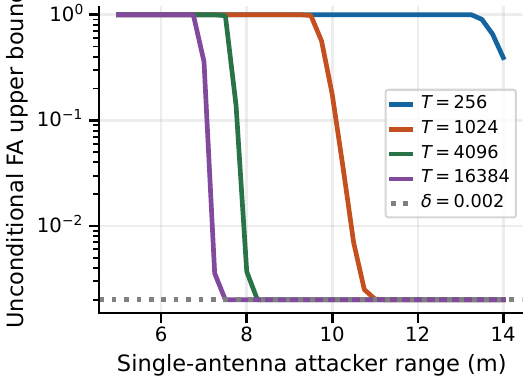}
		\label{fig:cost:a}%
	}
	\hfill
	\subfloat[]{%
		\includegraphics[width=0.47\columnwidth]
		{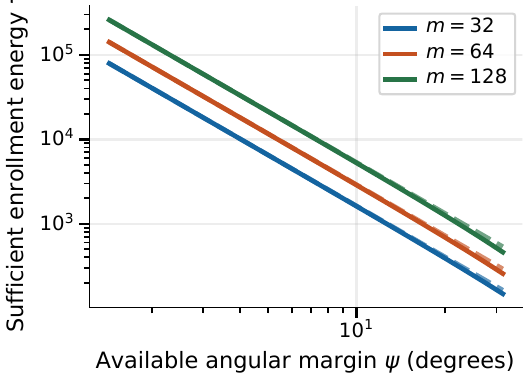}
		\label{fig:cost:b}%
	}
	\caption{Pre-enrollment FA bounds and the energy needed for angular separation.
		(a) Bounds for a template-aware single-antenna attacker.
		Blue, orange, green, and purple denote total enrollment
		energies of $T=256$, $1024$, $4096$, and $16384$. Solid
		curves use \eqref{eq:envelope} and \eqref{eq:outagecert}.
		The dotted gray line marks the enrollment-outage budget
		$\delta=0.002$. The bounds apply at the sampled geometries
		even when the attacker chooses its precoder after enrollment.
		(b) Sufficient received enrollment energy versus the
		angular margin $\psi$ in \eqref{eq:margin} for
		$\delta=0.002$. Blue, orange, and green denote $m=32$,
		$64$, and $128$, respectively. Solid curves come from
		numerical inversion of \eqref{eq:trainingbudget}.
		Dashed curves give the inverse-square approximation in
		\eqref{eq:costlaw}. The vertical axis gives energy
		sufficient for the angular-envelope condition. Both
		panels report analytical results evaluated numerically.}
	\label{fig:cost}
\end{figure}

\subsection{Certification Before Enrollment and Acquisition Cost}
Figure~\ref{fig:cost}(a) applies Corollary~\ref{cor:unconditional}
to a single-antenna attacker at the sampled broadside ranges.
The bound is computed before enrollment and covers attacks
chosen after the template is observed. Its floor at $\delta$
comes from assigning an acceptance bound of one to enrollments
outside $\mathcal G$. It is an outage allowance in the
certificate rather than an empirical FA level. At $5$~m, all four bounds equal one and give no useful limit on FA.
With $T=256$, this also holds over much of the sampled range.
Its spatial scope is the stated channel model and sampled
geometries. A continuous two-dimensional exclusion claim
would need a bound over the entire region.

Figure~\ref{fig:cost}(b) compares numerical inversion of
\eqref{eq:trainingbudget} with the approximation in
\eqref{eq:costlaw}. The inverse-square law becomes accurate
as the angular margin shrinks. The plotted quantity is the
received enrollment energy sufficient for the angular
envelope. Converting it to a pilot count requires the
per-pilot received SNR. A larger regional dimension
increases the Gamma quantile and thus the required energy
for this condition.

\section{Conclusion}

We studied how noisy enrollment affects near-field PLA. We showed that a fixed residual-energy threshold can cause
the FR probability to approach one at high received energy.
The proposed test scales the allowable residual energy with the aligned
energy. Its calibration controls FR for arbitrary regional
gains and all legitimate received energies above a prescribed minimum.
For an attacker that knows the enrolled reference, the attainable-energy
frontier characterizes the maximum acceptance probability. The
pre-enrollment bound accounts for randomness in the enrolled reference
and allows the attacker to adapt its precoder to the observed reference.
For a single region and a fixed angular threshold, halving a small angular
margin requires approximately four times the enrollment energy to satisfy
the same angular condition.
In the numerical study, the FR upper bound was $0.9984\%$.
Among the tested partitions, $J=4$ provided the best FR--FA
trade-off. At a regional phase perturbation of $20^\circ$, its
simulated FR was $0.1442\%$, while its FA at $9$~m was $1.084\%$
under the fixed transmit-energy setting.
For the enrolled reference used in the
three-antenna attack experiment at $8$~m, the numerically attained acceptance
probability was approximately $2.22\times10^{-5}$, and the corresponding
Chernoff upper bound was approximately $2.34\times10^{-4}$.

	\IfFileExists{references.bib}{
		\bibliographystyle{IEEEtran}
		\bibliography{references}
	}{
	}
\end{document}